\documentclass[12pt,english]{article}
\usepackage[T1]{fontenc}
\usepackage[latin9]{inputenc}
\usepackage{geometry}
\usepackage{caption}
\usepackage{subcaption}
\usepackage{color}
\usepackage{babel}
\usepackage{array}
\usepackage{tikz}
\usetikzlibrary{intersections}
\usetikzlibrary{positioning}
\usepackage{comment}
\usepackage{float}
\usepackage{tikz}
\usepackage{amsmath}
\usepackage{amsthm}
\usepackage{amssymb}
\usepackage{graphicx}  
\usepackage{setspace}
\usepackage{esint}
\usepackage[authoryear]{natbib}
\usepackage{pgfplots}
\pgfplotsset{compat=1.18}
\makeatletter
\usepackage{multirow}
\usepackage{makecell}
\usepackage{threeparttable}

\usepackage{babel}

\newtheorem{notation}{Notation}

\makeatother

\begin{document}
%
\title{Manipulation testing based on Benford's Law for discrete scores}

\author{Roy Cerqueti\thanks{Department of Social and Economic Sciences, Sapienza University of Rome, Italy, and GRANEM, University of Angers, France.
Email: {roy.cerqueti@uniroma1.it}},
Marco Ventura\thanks{Department of Economics and Law, Sapienza University of Rome, Italy. Email:
{marco.ventura@uniroma1.it}. \\ The authors wish to thank the audience of the 2025 IAEE Conference and the participants at the DiSSE Sapienza and HEC UNIL seminars for their constructive feedback. We are also indebted to M. Cano-Rodriguez for his support and thoughtful suggestions at an early stage of the project.}}

\maketitle
\begin{abstract}
\begin{singlespace}
This paper addresses the problem of running variable manipulation in Regression Discontinuity Designs. Leveraging the observation that manipulation often alters the density balance around the cutoff, we detect these structural imbalances using Benford's Law -a natural statistical regularity widely applied in fraud detection. Our framework serves as a vital precautionary safeguard alongside traditional McCrary-type tests. It eliminates researcher-chosen parameters that can skew outcomes, while delivering a deeper diagnostic breakdown of the density's behavior. Crucially, whereas the classic McCrary test can overlook systemic imbalances due to its rigid symmetric setup, our method separates the data into directional components. This allows researchers to pinpoint the exact origin of a deviation and spot hidden manipulation that standard frameworks fail to capture. To achieve this, we introduce an innovative method for selecting a bandwidth consistent with BL, and construct two distinct, complementary tests using threshold values adapted from \cite{nigrini} that successfully transition the law's application from digits to probabilities. Empirical applications confirm the enhanced protective value of this diagnostic framework.
\bigskip

\textbf{Keywords:} Regression discontinuity design; Benford's Law; Optimal bandwidth; Manipulation test \\
\end{singlespace}
\textbf{JEL Classification}: C01, C12, C21, C46

\newpage
\end{abstract}

\section{Introduction} \label{sec:intro}
Regression discontinuity design (RDD) is a widely utilized treatment effect estimator and is considered one of the most credible techniques. Like any other estimator, its validity hinges on a set of identifying assumptions. A crucial assumption is the absence of sorting or manipulation, which posits that units do not have perfect control over the running variable (or score) used to determine treatment. This assumption ensures that the assignment to treatment is effectively ``as if random'' around the cutoff point. In some empirical applications, the presence or absence of manipulation can be readily assessed based on the study design. For example, when the running variable is temporal, such as individuals' or firms' age, and the cutoff is defined as a specific point in time (e.g., age for eligibility for the treatment), manipulation is unlikely if the events influencing the outcome occurred prior to the cutoff. A case like this is reported by \cite{mellace2023short}, who analyze a policy fostering young innovative firms in Italy that can be considered as eligible based on their date of birth. In many other cases, manipulation is a potential concern, and its likelihood must be empirically assessed. \\
When perfect manipulation is not possible, random change would place roughly the same number of units on either side of the cutoff, leading to a continuous probability density function when the score is continuously distributed. Along this line, the seminal work by \cite{mccrary2008manipulation} introduced the idea to estimate twice the density of the units around the cutoff once at the left-hand-side (LHS) and once at the right-hand-side (RHS) and to perform an equality test. From a statistical point of view, the idea was first implemented by a nonparametric local-polynomial estimator, which introduces additional tuning parameters to pre-bin the data. Successively, the methodology was enhanced by \cite{otsu2013estimation}, who used boundary-corrected kernels, and more recently \cite{cattaneo2020simple} developed a set of manipulation tests based on local-polynomial density estimator, which does not require pre-binning of the data. \\ The null hypothesis of these McCrary-type tests is that there is ``no manipulation" of the density at the cutoff, formally stated as continuity of the density functions for control and treatment units at the cutoff. It follows that failing to reject the null implies no statistical evidence of manipulation and offers evidence supporting the validity of the RDD \citep{cattaneo2019practical}.\\
The papers mentioned are of paramount relevance in the RDD environment. However, they offer a method for assessing manipulation only when the score has a continuous distribution. In most of the applications, assuming the continuity of the running variable is hard and sometimes inadequate. By contrast, in empirical cases a discrete setting is standard, where the score takes a countable and finite number of realizations. It follows that one cannot use the concept of continuity and must refer to the weaker concept of asymmetry around the cutoff. 
\\
In this vein, an equivalent version of the McCrary-type test for discrete running variables is given by \cite{frandsen2017party}, who follows the same idea, counting units on the two sides of the cutoff. This approach is reported also in \cite{cattaneo2024practical}, where the authors present the density of the running variable by discussing a falsification test counting the number of observations above and below the cutoff (see Section 2.3.2). However, a counting procedure does not guarantee a difference between the two sides of the cutoff. One can reproduce the same behavior of the score above and below the cutoff, even if the number of occurrences varies.
\\
This paper elaborates on this point. Specifically, we propose a new type of test for manipulation of discrete scores -- i.e., scores taking a finite number of values --  that is realistically grounded in the cutoff-based symmetry of the density instead of unreliable concepts of continuity and counting procedures. Specifically, we adopt an approach based on a natural law, notably the so-called \textit{Benford's Law} (BL). As we will see, such a law is particularly suitable for our purposes. One of the main advantage of our strategy is to avoid discretionary choices on the part of the researcher. In particular, when carrying out a McCrary-type test following the procedure suggested by \cite{cattaneo2018manipulation}, the researcher selects the local polynomial order to construct the density estimators, the local polynomial order used to construct the bias-corrected density estimators, the density estimator method and the kernel function. In contrast, BL provides us with predetermined thresholds of the support of the running variable within which we carry out the comparison between RHS and LHS of the cutoff. Furthermore, BL predetermines the density values for these thresholds, thus avoiding estimation issues. Importantly, BL is not an \textit{ad-hoc} choice. Such a probabilistic law is a natural property of the digits of the elements of empirical datasets; thus, it is a universal regularity of the data that removes possible arbitrariness in the analytical approach.  For an overview of BL, we refer to \cite{berger2015}. The main characteristics of this statistical law are appreciated mainly when BL does not hold. Indeed, this case is often associated with data manipulations and artificial intervention to modify the elements of the dataset. Literature provides clear evidence on the matter \citep[see e.g.,][]{arezzo2023benford, ausloos2016, ausloos2021, mir2014, cerioli2019, todter2009}. For this reason, BL is particularly appropriate in our context. However, this natural law tends to fail in some cases that are inherently associated with the constitutive properties of the dataset considered. Specifically, the dataset should be of large cardinality, and data should not be theoretically bounded in a finite range.
Importantly, as we will see, this paper overcomes these limitations by referring merely to the statistical definition of the BL as a discrete probability distribution. Indeed, we avoid the constraint of referring exclusively to the digits of the data, hence corroborating the universality of the approach followed. It follows that another advantage of our procedure is that the bandwidth (BW) within which carrying out the test is automatically suggested by BL, avoiding once again discretionary choices of the optimum data-driven method to be chosen.
\\
The procedure presents two phases. In the first phase, we provide a new BL-based optimization criterion for computing an optimal data-driven BW around the cutoff; this criterion is of the iterative type. In the second phase, we propose two new methods for assessing the asymmetry of the running variable at the cutoff. Briefly, the procedure for assessing the asymmetry is as follows. We split the distribution of the running variable within the optimal BW into two parts -- above and below the cutoff. Then, we adopt a twofold approach. In the first approach, we identify the thresholds on the $x$-axis such that the running variable obeys the BL probability distribution. Such an identification is implemented above and below the cutoff, after a suitable rescaling of these two parts of the density function such that their integral on the two sides sums up to unity. Then, these thresholds are statistically compared to determine whether the distribution exhibits symmetry properties when observed before and after the cutoff. In the second approach, we use the thresholds found above for deriving the ones having the same distance from the cutoff and compare the empirical probabilities through a MAD test. These approaches provide a comprehensive view of all aspects related to the asymmetry in the case of discrete running variables. \\


The paper proceeds as follows. Section \ref{sec:benford} introduces BL and its properties, along with its applications in economics. Section \ref{sec:optimal_BW} explains the iterative procedure for deriving the optimal BW. Section \ref{sec:testing} sets out the testing methodologies. Section \ref{sec:simulation} contains extensive simulations, while two applications to relevant empirical cases is offered in Section \ref{sec:applications}. Finally, Section \ref{sec:conclusion} presents some concluding remarks.

\section{A brief introduction to Benford's Law and its applications in economics} \label{sec:benford}
The BL belongs to the natural law category, arising from observations of real-world phenomena. Indeed, it was first observed by Simon Newcomb in 1881 \citep{necomb1881note}, who noticed the pages of logarithm tables near the beginning of the book were more worn. Frank Benford independently rediscovered and popularized the law in 1938 \citep{benford1938law}.
BL, also known as the First-Digit Law, describes the surprising non-uniform distribution of leading digits in many real-world datasets. Contrary to the intuitive expectation of equal probability for each digit $1, \dots, 9$, BL predicts a highly skewed distribution, with smaller digits appearing more frequently than larger ones. Specifically, a set of numbers is said to satisfy BL if the leading digit $j$, for $j \in \{1, \dots,9 \}$, occurs with probability

\begin{equation} \label{eq:definition}
    B_j = 
    \log_{10} \left(1+\frac{1}{j} \right) .
\end{equation}

\noindent The quantity $B_j$ is proportional to the ($\log$ of the) space between $j$ and $j+1$. It follows that this is the distribution expected if the log of the numbers are uniformly and randomly distributed. Since the interval $\left[\log_{10}(1), \log_{10}(2)\right]$ is wider than the interval $\left[\log_{10}(8), \log_{10}(9)\right]$ (around $0.30$ and $0.05$, respectively) a randomly chosen value from a uniform distribution on the log scale is more likely to fall within the wider interval, making numbers starting with $1$ more likely than those starting with $9$. Notice also that \eqref{eq:definition} still applies also in other $\log$ basis. \\
\indent Some key contributing factors help explaining this phenomenon, such as: $(i)$ scale invariance: the law remains largely unaffected by changes in units of measurement, suggesting its origin lies in underlying multiplicative processes. $(ii)$ Invariance under power laws: many natural and social phenomena exhibit power-law distributions (e.g., city sizes, earthquake magnitudes, income distribution). These lead to skewed leading digit distributions. $(iii)$ Multiplicative growth processes: economic and financial data often involve multiplicative growth (e.g., compound interest, economic growth), which naturally generate data conforming to BL. \\
The law has had disparate applications, such as genomics data, election data, and criminal cases, to the extent that in the US, evidence based on BL has been admitted in criminal cases at all levels of the court system (for applications of BL, see \cite{berger2015} and the references therein). Also in economics, the law has already raised attention. \cite{varian26benford} first noticed that the law could be used to detect possible fraud in lists of socio-economic data submitted in support of public planning decisions. Based on the assumption that people who fabricate figures tend to distribute their digits fairly uniformly, a simple comparison of first-digit frequency distribution from the data with the expected distribution according to BL ought to show up any anomalous results. A similar reasoning has been applied to accounting data \citep{arezzo2023benford}, pricing patterns \citep{el2005price}, scientific data validation \citep{diekmann2007not}, macroeconomic data report \citep{rauch2011fact}, and network analysis \citep{tovsic2021use}. In this article, we follow the intuition by \cite{varian26benford} and propose a manipulation test based on BL in a RDD context. \\ 
The BL accuracy has been documented by \cite{cai2020surprising}, but despite its attractiveness, it does not apply universally. Short datasets, artificially generated data, and data with specific constraints may not exhibit the expected digit distribution. However, in our approach, we do not refer to digits, but rather to the thresholds generated by the BL and the ensuing distribution, which allow to sidestep these limitations.

\section{The optimal BL bandwidth}
\label{sec:optimal_BW}
Due to obvious endogeneity problems, the testing procedure must be executed within an optimal BW. In our case, the optimality criterion should be refined consistently with the context we are working with. So that, in what follows, we proceed to derive a new data-driven optimality criterion based on the BL. 
Notably, the optimal BW is found through an iterative procedure that solves an optimization problem for determining BL-based thresholds for the running variable.

\subsection{The iterative procedure}
\label{sec:BW}
 The running variable $R_i$ is the quantity of interest. We denote the probability density function of the running variable $R_i$ by $f:\mathbb{R} \to \mathbb{R}$ and the cutoff by $C \in \mathbb{R}$. We deal with the potential asymmetry of $f$ above and below the cutoff, so that $f$ exhibits different behaviors to the left and to the right of $C$.

\begin{itemize}
\item We start by considering the support of $R_i$, namely $g(R_i) \subseteq \mathbb{R}$.
\\
Specifically, we define 
$$g^{(-)}(R_i)=\{x \in g(R_i) : x \leq C\}, \qquad g^{(+)}(R_i)=\{x \in g(R_i) : x > C\}.$$ By construction, $g^{(-)}(R_i) \cap g^{(+)}(R_i)= \emptyset$ and 
$g^{(-)}(R_i) \cup g^{(+)}(R_i)=g(R_i)$.
\\
We then select two sets of endogenous thresholds using BL: $\mathcal{X}_1^{(-)}=\{X^{(-)}_{1,1}, \dots, X^{(-)}_{1,8}\} \subset \mathbb{R}$ and $\mathcal{X}_1^{(+)}=\{X^{(+)}_{1,1}, \dots, X^{(+)}_{1,8}\} \subset \mathbb{R}$ such that
$$X^{(-)}_{1,8} < \dots < X^{(-)}_{1,1} < C < X^{(+)}_{1,1} < \dots< X^{(+)}_{1,8}$$ and satisfying the following property:
$$\sum_{x \in g(R_i)} f(x)\mathbf{1}(x \in [a,b))=$$
\begin{equation}
\label{eq:prop1}
= \begin{cases}
\gamma_1 B_1 & \text { if } (a,b)=(C,X^{(+)}_{1,1}) \text{ or } (a,b)=(X^{(-)}_{1,1},C), \\
\gamma_1 B_j & \text { if } (a,b)=(X^{(+)}_{1,j-1},X^{(+)}_{1,j}) \text{ or } (a,b)=(X^{(-)}_{1,j},X^{(-)}_{1,j-1}), \text{ for } j=2, \dots 8,\\
\gamma_1 B_9 & \text { if } (a,b)=(X^{(+)}_{1,8},+\infty) \text{ or } (a,b)=(-\infty,X^{(-)}_{1,8}),
\end{cases} 
\end{equation}
where $\gamma_1$ is a normalization constant given by
\begin{equation}
\label{eq:propgamma1}
\gamma_1= \begin{cases}
\sum_{x \in g^{(-)}(R_i)} f(x) & \text { when using the thresholds in } \mathcal{X}_1^{(-)}  \\
\mbox{} \\
\sum_{x \in g^{(+)}(R_i)} f(x) & \text { when using the thresholds in } \mathcal{X}_1^{(+)},
\end{cases} 
\end{equation}
where 
$\mathbf{B}=(B_1, \dots, B_9)$ is the first digit-Benford distribution, having components defined as in \eqref{eq:definition}. 

\newtheorem*{remark}{Remark}
\begin{remark}
This point is crucial because it clarifies how BL operates here. Indeed, we do not assume $R$ to obey the BL; that is, we do not assume that the digits of the $R$ variable are distributed according to $P(R=j)=\log_{10}\left(1+\frac{1}{j} \right) $. Conversely, we are partitioning the $R$ range according to the probabilities consistent with the BL, instead of using a different law, such as a uniform, triangular, etc., distribution. The rationale underpinning this choice is that the BL is a natural law, unlike the uniform, triangular, or any other probability density function. As a result, one does not need $R$ to obey the BL, and the procedure is sufficiently general to be implemented in any context, irrespective of whether BL applies to the digits of $R$.
\end{remark}

\item In the second step, we restrict to $[C-X^{(-)}_{1,1}, C]$ and $[C,C+X^{(+)}_{1,1}]$ by exploiting BL by adapting the procedure presented for the first step.
\\
We then find two sets of endogenous thresholds $\mathcal{X}_2^{(-)}=\{X^{(-)}_{2,1}, \dots, X^{(-)}_{2,8}\} \subset \mathbb{R}$ and $\mathcal{X}_2^{(+)}=\{X^{(+)}_{2,1}, \dots, X^{(+)}_{2,8}\} \subset \mathbb{R}$ such that
$$X^{(-)}_{2,8} < \dots < X^{(-)}_{2,1} < C < X^{(+)}_{2,1} < \dots< X^{(+)}_{2,8}$$ and satisfying the following property:
$$\sum_{x \in g(R_i)} f(x)\mathbf{1}(x \in [a,b))=$$
\begin{equation}
\label{eq:prop2}
= \begin{cases}
\gamma_2 B_1 & \text { if } (a,b)=(C,X^{(+)}_{2,1}) \text{ or } (a,b)=(X^{(-)}_{2,1},C), \\
\gamma_2 B_j & \text { if } (a,b)=(X^{(+)}_{2,j-1},X^{(+)}_{2,j}) \text{ or } (a,b)=(X^{(-)}_{2,j},X^{(-)}_{2,j-1}), \text{ for } j=2, \dots 8,\\
\gamma_2 B_9 & \text { if } (a,b)=(X^{(+)}_{2,8},X^{(+)}_{1,1}) \text{ or } (a,b)=(X^{(-)}_{1,1},X^{(-)}_{2,8}),
\end{cases} 
\end{equation}
where $\gamma_2$ is a normalization constant given by
\begin{equation}
\label{eq:propgamma2}
\gamma_2= \begin{cases}
\sum_{x \in g(R_i)} f(x)\mathbf{1}(x \in [X^{(-)}_{1,1},C)) 
 & \text { when using the thresholds in } \mathcal{X}_2^{(-)}  \\
\mbox{} \\
\sum_{x \in g(R_i)} f(x)\mathbf{1}(x \in [C,X^{(+)}_{1,1}))
 & \text { when using the thresholds in } \mathcal{X}_2^{(+)}.
\end{cases} 
\end{equation}
\item In the generic $n$-th step, with $n =3,4, \dots$, we consider $[C-X^{(-)}_{n-1,1}, C]$ and $[C,C+X^{(+)}_{n-1,1}]$ as in the previous steps and find two sets of endogenous thresholds $\mathcal{X}_{n}^{(-)}=\{X^{(-)}_{n,1}, \dots, X^{(-)}_{n,8}\} \subset \mathbb{R}$ and $\mathcal{X}_{n}^{(+)}=\{X^{(+)}_{n,1}, \dots, X^{(+)}_{n,8}\} \subset \mathbb{R}$ such that
$$X^{(-)}_{n,8} < \dots < X^{(-)}_{n,1} < C < X^{(+)}_{n,1} < \dots< X^{(+)}_{n,8}$$ and satisfying the following property:
$$\sum_{x \in g(R_i)} f(x)\mathbf{1}(x \in [a,b))=$$
\begin{equation}
\label{eq:propn}
= \begin{cases}
\gamma_{n} B_1 & \text { if } (a,b)=(C,X^{(+)}_{n,1}) \text{ or } (a,b)=(X^{(-)}_{n,1},C), \\
\gamma_{n} B_j & \text { if } (a,b)=(X^{(+)}_{n,j-1},X^{(+)}_{n,j}) \text{ or } (a,b)=(X^{(-)}_{n,j},X^{(-)}_{n,j-1}), \text{ for } j=2, \dots 8,\\
\gamma_{n} B_9 & \text { if } (a,b)=(X^{(+)}_{n,8},X^{(+)}_{n-1,1}) \text{ or } (a,b)=(X^{(-)}_{n-1,1},X^{(-)}_{n,8}),
\end{cases} 
\end{equation}
where $\gamma_{n}$ is a normalization constant given by
\begin{equation}
\label{eq:propgamman}
\gamma_n= \begin{cases}
\sum_{x \in g(R_i)} f(x)\mathbf{1}(x \in [X^{(-)}_{n-1,1},C)) & \text { when using the thresholds in } \mathcal{X}_{n-1}^{(-)}  \\
\mbox{} \\
\sum_{x \in g(R_i)} f(x)\mathbf{1}(x \in [C,X^{(+)}_{n-1,1}))& \text { when using the thresholds in } \mathcal{X}_{n-1}^{(+)}.
\end{cases} 
\end{equation}
\begin{notation}
    For the sake of simplicity, henceforth we will say that $(a,b)$ is as in Formula \eqref{eq:propn} to say that it is defined as in the RHS of \eqref{eq:propn} for $j=1, \dots, 9$.
    \\
    Furthermore, when needed, we use the equivalence stating that $X^{(+)}_{n,9}=X^{(+)}_{n-1,1}$ and $X^{(-)}_{n,9}=X^{(-)}_{n-1,1}$ 
\end{notation}

\item The last step runs at $n=N$, where $N\in \mathbb{N}$ is fixed at the beginning. 
\end{itemize}

It is important to notice that the number of available data points might affect the approximation of Benford's distribution with the thresholds in  $\mathcal{X}_{n}^{(-)}$ and $ \mathcal{X}_{n}^{(+)}$, with $n=1, \dots, N$. In fact, as the number of data points increases, the computed thresholds yield a more accurate approximation of Benford's distribution. Conversely, the approximation deteriorates as the number of steps $n$ increases because the number of available observations tends to decrease. \\ On the one hand, this consideration seems to suggest that a limited number of iterations may be desirable to have a good approximation.
On the other hand, we need to be as close as possible to the cutoff $C$ to reduce bias and to check for the asymmetry of the density function $f$. Therefore, we have a counteracting force pushing towards a high number of steps $n$. Substantially, there exists a trade-off between these two competing forces.\\
\indent To describe the way the optimal BW is selected, we need an empirical dataset. We assume that this dataset has observed values of the running variable given by the set
$$
\mathcal{R}= \{r_1, \dots, r_K\}.
$$
We assume that this empirical dataset has the function $f$ used in the steps above as its empirical probability density function.
Using the theoretical thresholds suggested by BL, $\mathcal{X}_{n}^{(-)}$ and $ \mathcal{X}_{n}^{(+)}$, we compute the empirical distribution, i.e., the histograms, by setting
\begin{equation}
\label{eq:Btilde}
\tilde{B}^{(-)}_{n,j}= \frac{1}{R_n} \sum_{k=1}^K r_k \mathbf{1}(r_k \in [a,b))\,\,\,\,\, \text{  and  } \,\,\,\,\,
\tilde{B}^{(+)}_{n,j}= \frac{1}{R_n} \sum_{k=1}^K r_k \mathbf{1}(r_k \in [a,b)),
\end{equation}
where $(a,b)$ is as in Formula \eqref{eq:propn}, $j=1, \dots, 9$, $n=1, \dots, N$ and
\begin{itemize}
\item For $n=1$
\begin{equation}
\label{R_n}
R_n= \begin{cases}
\sum_{k=1}^K r_k  \mathbf{1}(r_k \in (-\infty,C)) & \text { when using the thresholds in } \mathcal{X}_{1}^{(-)},\\
\mbox{} \\
\sum_{k=1}^K r_k  \mathbf{1}(r_k \in [C,+\infty)) & \text { when using the thresholds in } \mathcal{X}_{1}^{(+)} ;
\end{cases} 
\end{equation}
\item for $n=2, \dots, N$
\begin{equation}
\label{R_n}
R_n= \begin{cases}
\sum_{k=1}^K r_k  \mathbf{1}(r_k \in (X^{(-)}_{n-1,1},C)) & \text { when using the thresholds in } \mathcal{X}_{n-1}^{(-)}, \\
\mbox{} \\
\sum_{k=1}^K r_k  \mathbf{1}(r_k \in [C,X^{(+)}_{n-1,1})) & \text { when using the thresholds in } \mathcal{X}_{n-1}^{(+)}.
\end{cases} 
\end{equation}
\end{itemize}
In words, $\tilde{B}^{(-)}_{n,j}$ and $\tilde{B}^{(+)}_{n,j}$ are the share of data points in the interval $[a,b)$ suggested by the BL. 
The idea to optimally select the BW consists in finding $n^\star \in \{1, \dots, N\}$ such that the couple $(X^{(-)}_{n^\star,1}, X^{(+)}_{n^\star,1})$ leads to a compliance of the empirical probability distributions in \eqref{eq:Btilde} with the BL according to a distance criterion. In our context, we exploit the MAD, that is particularly suitable in the BL context \citep[see e.g.,][]{cerqueti2021data}. Specifically,
$$
MAD_n^{(-)}= \frac{1}{9} \sum_{j=1}^9|\tilde{B}^{(-)}_{n,j}-B_j| \,\,\,\,\, \text{  and  } \,\,\,\,\, MAD_n^{(+)}=\frac{1}{9} \sum_{j=1}^9|\tilde{B}^{(+)}_{n,j}-B_j|.
$$
The $MAD$ criterion can be regarded as the mean of the absolute deviations between the empirical and the theoretical BL distribution, i.e., a summary statistic of the distance between the observed distribution and the one predicted by the BL. 
According to the arguments above, the $MAD$ tends to increase as $n$ increases, but we need $n$ sufficiently large to compute many well-accurate differences. Therefore, we can introduce an acceptance level $\alpha$ for $MAD$ and define the $\alpha$-optimal BW $n_\alpha^\star$ by setting 
\begin{equation}
\label{nstar}
n_\alpha^\star  = \sup\left\{n=1, \dots, N \,: \, MAD_{n-1}^{(-)} < \alpha \text{ and } MAD_{n-1}^{(+)} < \alpha \right\}.
\end{equation}
The $\alpha$-optimal couple is $(X^{(-)}_{n_\alpha^\star,1}, X^{(+)}_{n_\alpha^\star,1})$. For an easy notation, we set hereafter $X^{(-)}_{\alpha,\star}:=X^{(-)}_{n_\alpha^\star,1}$ and $ X^{(+)}_{\alpha,\star}:=X^{(+)}_{n_\alpha^\star,1}$\\

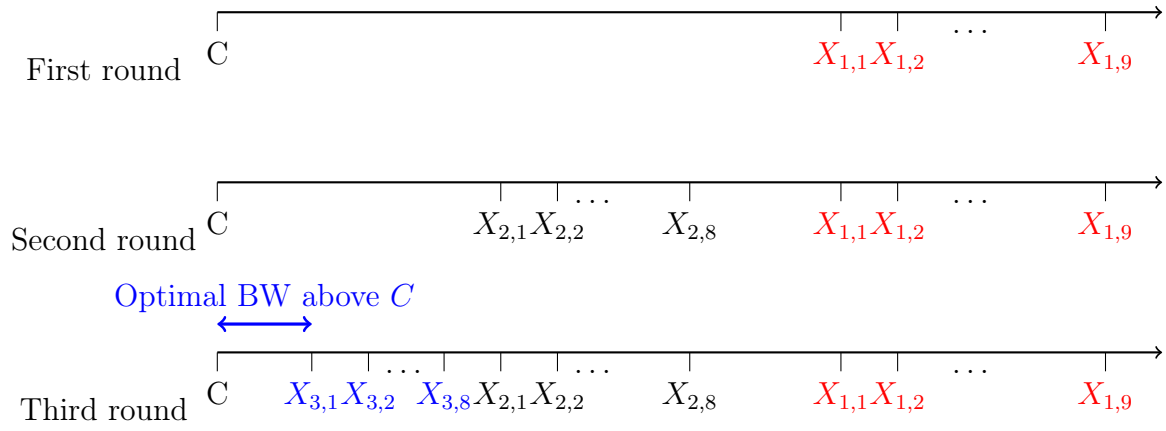
\begin{figure}[ht]
    \centering
    \begin{tikzpicture}[scale=0.25] 

 \begin{scope}[yshift=18cm]
\node at (-1,-3) {First round};
\draw[thick,->] (5,0) -- (55,0);
\draw (5,0) -- (5,-1) node[below] {C};
\draw (38,0) -- (38,-1) node[below] {\textcolor{red}{$X_{1,1}$}};
\draw (41,0) -- (41,-1) node[below] {\textcolor{red}{$X_{1,2}$}};
\node at (45,-1) {$\dots$};
\draw (52,0) -- (52,-1) node[below] {\textcolor{red}{$X_{1,9}$}};
\end{scope}
        
   \begin{scope}[yshift=9cm]
     \node at (-1,-3) {Second round};
     \draw[thick,->] (5,0) -- (55,0);
     \draw (5,0) -- (5,-1) node[below] {C};
     \draw (20,0) -- (20,-1) node[below] {$X_{2,1}$};
     \draw (23,0) -- (23,-1) node[below] {$X_{2,2}$};
     \node at (25,-1) {$\dots$};        
     \draw (30,0) -- (30,-1) node[below] {$X_{2,8}$};
     \draw (38,0) -- (38,-1) node[below] {\textcolor{red}{$X_{1,1}$}};
     \draw (41,0) -- (41,-1) node[below] {\textcolor{red}{$X_{1,2}$}};
     \node at (45,-1) {$\dots$};
     \draw (52,0) -- (52,-1) node[below] {\textcolor{red}{$X_{1,9}$}};
     \end{scope}
        
    \node at (-1,-3) {Third round};
    \draw[thick,->] (5,0) -- (55,0); 
    \draw (5,0) -- (5,-1) node[below] {C}; 
    \draw (10,0) -- (10,-1) node[below] {\textcolor{blue}{$X_{3,1}$}};
    \draw (13,0) -- (13,-1) node[below] {\textcolor{blue}{$X_{3,2}$}};
    \node at (15,-1) {$\dots$};
    \draw (17,0) -- (17,-1) node[below] {\textcolor{blue}{$X_{3,8}$}};
    \draw (20,0) -- (20,-1) node[below] {$X_{2,1}$};
    \draw (23,0) -- (23,-1) node[below] {$X_{2,2}$};
    \node at (25,-1) {$\dots$};        
    \draw (30,0) -- (30,-1) node[below] {$X_{2,8}$};
    \draw (38,0) -- (38,-1) node[below] {\textcolor{red}{$X_{1,1}$}};
    \draw (41,0) -- (41,-1) node[below] {\textcolor{red}{$X_{1,2}$}};
    \node at (45,-1) {$\dots$};
    \draw (52,0) -- (52,-1) node[below] {\textcolor{red}{$X_{1,9}$}};

    \draw[very thick, blue, <->] (5,1.5) -- (10,1.5) node[midway, above] {\textcolor{blue}{Optimal BW above $C$}};

\end{tikzpicture}
    \caption{Iteration for deriving the optimal BW in the case above the cutoff $C$.}
    \label{fig:graph_BW} 
\end{figure}

Notice that the trivial case where no $n^\star \in \{1, \dots, N\}$ exists such that the pair $(X^{(-)}_{n^\star,1}, X^{(+)}_{n^\star,1})$ complies with BL typically arises when the dataset contains a very limited number of distinct values. Our procedure excludes these uninformative instances. Figure \ref{fig:graph_BW} provides a visual representation of the procedure shown in Section \ref{sec:BW} for the case of $g^{(+)}(R_i)$ and assuming that $n^\star=3$.

\section{Testing for asymmetry}
\label{sec:testing}
Comparing the thresholds in $\mathcal{X}^{(-)}$ and $\mathcal{X}^{(+)}$ is a way of assessing the symmetry features of the running variable. This is exactly the point where our approach differs from those based on density comparison; we use threshold values. \\ Operationally, we follow two different perspectives to face the symmetry problem, recalling that the methods proposed are applied within the BW found in the previous section. For the sake of simplicity, we refer to $X^{(-)}_{n,j}$ and $X^{(+)}_{n,j}$ simply as $X^{(-)}_{j}$ and $X^{(+)}_{j}$, respectively, for each $j=1, \dots, 9$.
\subsection{First method: the tolerance threshold approach}
\label{sec:method1}
In the case of a symmetric density function $f$, one has that 
\begin{equation}
\label{eq:cond}
C-X^{(-)}_j=X^{(+)}_j-C, \qquad \forall \, j=1, \dots, 9, 
\end{equation}
that can be rewritten as 
\begin{equation}
\label{eq:cond2}
2C-X^{(-)}_j-X^{(+)}_j=0, \qquad \forall \, j=1, \dots, 9.
\end{equation}
Reasonably, the condition of perfect equivalence between all the thresholds in $\mathcal{X}^{(-)}$ and $\mathcal{X}^{(+)}$ as in \eqref{eq:cond2} might be too restrictive, and defining a tolerance level may lead to subjective choices. To circumvent the problem we convert the discrepancies in \eqref{eq:cond2} into probabilities, whence the possibility to use the MAD and its associated critical values. 

We first set a threshold $\alpha>0$ -- which is the closely acceptance threshold for MAD by Nigrini -- set $X^{(-)}_9:=X^{(-)}_{\alpha,\star}$ and $X^{(+)}_9:=X^{(+)}_{\alpha,\star}$ and define $\Gamma=\max \{C-X^{(-)}_9, X^{(+)}_9-C \}$ that will be used to normalize the thresholds as:
$$
\tilde{X}^{(-)}_j= \frac{C-{X}^{(-)}_j}{\Gamma},\,\,\,\,\,\,\tilde{X}^{(+)}_j= \frac{{X}^{(+)}_j-C}{\Gamma}, \qquad \forall \, j=1, \dots, 9.
$$
Now, the thresholds $\tilde{X}^{(-)}$'s and $\tilde{X}^{(+)}$'s are probabilities that are collected into two sets, $\tilde{\mathcal{X}}^{(-)}$ and $\tilde{\mathcal{X}}^{(+)}$, respectively. 
At this point we are in the position to apply a distance measure, $D$, between the probabilities in $\tilde{\mathcal{X}}^{(-)}$ and $\tilde{\mathcal{X}}^{(+)}$, as follows:

\begin{equation}
\label{eq:test}
D(\tilde{\mathcal{X}}^{(-)}, \tilde{\mathcal{X}}^{(+)})=\frac{1}{9} \sqrt{\sum_{j=1}^9 |\tilde{X}^{(-)}_j-\tilde{X}^{(+)}_j|}
\end{equation}

\subsection{Second method: the MAD approach}
\label{sec:method2}
In this second set up, we assume that the endogenous thresholds $\mathcal{X}^{(-)}$ and $\mathcal{X}^{(+)}$ can be conveniently used to define other thresholds to be used for the assessment of the asymmetry in the running variable.
We distinguish the cases of the right and left tail of the distribution, which are respectively the sets $g^{(+)}(R_i)$ and $g^{(-)}(R_i)$. Also in this case, we consider a threshold $\lambda>0$ and set $X^{(-)}_9:=X^{(-)}_{\lambda,\star}$ and $X^{(+)}_9:=X^{(+)}_{\lambda,\star}$.
\\
Specifically, we introduce the thresholds $Y^{(+)}_k$ and $Y^{(-)}_k$ such that they are symmetric values of the $X$'s with respect to the cutoff $C$, i.e.:
$$
C-Y^{(-)}_k=X^{(+)}_k-C,\,\,\,\,\,C-X^{(-)}_k=Y^{(+)}_k-C,
$$
for each $k=1, \dots, 9$. We collect the $Y$'s into two sets $\mathcal{Y}^{(-)}=\{Y^{(-)}_1, \dots, Y^{(-)}_9\} \subset \mathbb{R}$ and $\mathcal{Y}^{(+)}=\{Y^{(+)}_1, \dots, Y^{(+)}_9\} \subset \mathbb{R}$. 
We consider
\begin{equation}
\label{eq:prop+}
\sum_{x \in g(R_i)} f(x) \mathbf{1}(x \in [a,b))= \begin{cases}
\gamma p_1^{(+)} & \text { if } (a,b)=(C,Y^{(+)}_1), \\
\gamma p_j^{(+)} & \text { if } (a,b)=(Y^{(+)}_{j-1},Y^{(+)}_j), \text{ for } j=2, \dots 9
\end{cases} 
\end{equation}
and, analogously, 
\begin{equation}
\label{eq:prop-}
\sum_{x \in g(R_i)} f(x) \mathbf{1}(x \in [a,b))
= \begin{cases}
\gamma p_1^{(-)} & \text { if } (a,b)=(Y^{(-)}_1, C), \\
\gamma p_j^{(-)} & \text { if } (a,b)=(Y^{(-)}_{j},Y^{(+)}_{j-1}), \text{ for } j=2, \dots 9
\end{cases} 
\end{equation}
where $\gamma$ is defined in  \eqref{eq:propgamma1} and the $p^{(+)}$'s and $p^{(-)}$'s form two probability distributions.
Then, we apply the MAD criterion again to measure the deviation between the BL-type distributions in \eqref{eq:definition} and the ones in \eqref{eq:prop+} and \eqref{eq:prop-}. Specifically, 
\begin{equation} \label{eq:MADs}
    MAD^{(+)}=\frac{1}{9}\sum_{j=1}^9 | p_j^{(+)}- B_j|, \qquad MAD^{(-)}=\frac{1}{9}\sum_{j=1}^9 |p_j^{(-)}- B_j|.
\end{equation}

Accordingly, we classify the BL-type distributions as symmetric whenever $MAD^{(+)}<\lambda$ and $MAD^{(-)}<\lambda$ and as asymmetric whenever at least one of these inequalities is violated.
If $MAD^{(+)}$ and $MAD^{(-)}$ signal symmetry we conclude in favour of no manipulation. \\ 
In principle, it may happen that only one of the two MADs signals divergence, which still implies asymmetry. This apparent contradiction actually provides a further source of information. Ultimately, the final judgment rests upon the specific economic case under investigation. One can, for instance, consider the economic incentive that units have to place their score on the RHS or the LHS of $C$. Think of a policy in which units scoring above $C$ receive a desirable treatment. In this case, if $MAD^{(+)}<\lambda$ and $MAD^{(-)}\geq \lambda$, this suggests manipulation with no economic effects. Conversely, manipulation engenders a substantial effect when $MAD^{(+)} \geq \lambda$ and $MAD^{(-)}<\lambda$. \textit{Mutatis mutandis}, very similar reasoning applies when units have an incentive to place their score below the cutoff. This is a remarkable innovation with respect to the McCrary-type tests.
\\

\begin{figure}[ht]
    \centering
\begin{tikzpicture}[>=stealth, scale=1.5]
    \draw[->, thick] (0,0) -- (7,0);
    
    \draw[thick, name path=curva] (0,0.2) .. controls (1,2.5) and (3,3.0) .. (4.5,2.0) .. controls (5.5,1.3) and (6.5,0.7) .. (7,0.5);

    
    \path[name path=verticaleA] (1.6, 0) -- (1.6, 4);
    \draw[blue, ultra thick, name intersections={of=curva and verticaleA, by=P1}] (1.6, 0) -- (P1);
    \fill (P1) circle (1.5pt); 

    \path[name path=verticaleB] (2.25, 0) -- (2.25, 4);
    \begin{scope}[name intersections={of=curva and verticaleB, by=P2}]
        \draw[red, ultra thick] (2.25, 0) -- (2.25, 0.45);
        \draw[red, ultra thick] (2.25, 1.05) -- (P2);
        \fill (2.25, 0) circle (1.5pt);  
        \fill (P2) circle (1.5pt); 
    \end{scope}

    \path[name path=verticaleC] (3.6, 0) -- (3.6, 4);
    \draw[black, dashed, ultra thick, name intersections={of=curva and verticaleC, by=P3}] (3.6, 0) -- (P3);
    \fill (3.6, 0) circle (1.5pt); 

    \path[name path=verticaleD] (5.4, 0) -- (5.4, 4);
    \draw[red, ultra thick, name intersections={of=curva and verticaleD, by=P4}] (5.4, 0) -- (P4);
    \fill (5.4, 0) circle (1.5pt);    
    \fill (P4) circle (1.5pt); 

    \node[blue, below] at (1.6, -0.05) {\Large $Y_1^{(+)}$};
    \node[red, below] at (2.25, -0.05) {\Large $X_1^{(-)}$};
    \node[black, below] at (3.6, -0.05) {\Large $C$};
    \node[red, below] at (5.4, -0.05) {\Large $X_1^{(+)}$};

    \node[blue] at (2.25, 0.75) {\Large $p_1^{(+)}$};

    \node[red] at (2.9, 1.7) {\Huge $B_1$};
    \node[red] at (4.5, 1.1) {\Huge $B_1$};

\end{tikzpicture}
    \caption{Visual representation of the two testing procedures. The first method compares $X_1^{(+)}$ and $X_1^{(-)}$. The second method compares the areas below the curve and between $C$ and $X_1^{(-)}$ with the area between  $X_1^{(+)}$ and $C$.}
    \label{fig_flow}
\end{figure}
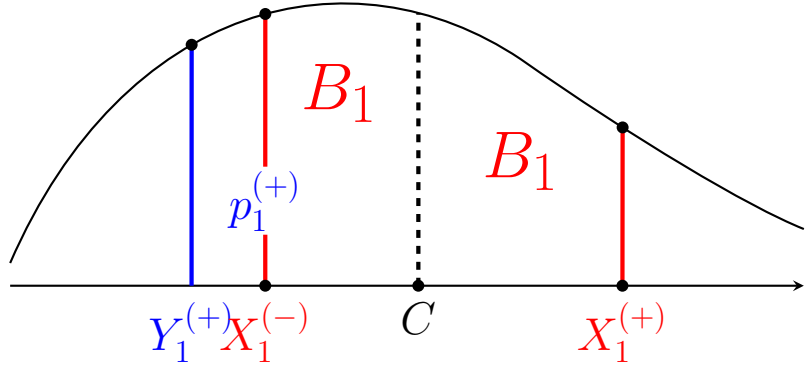

Figure \ref{fig_flow} visualizes the two testing procedures. The first method (shown in red) compares the thresholds $X_1^{(+)}$ and $X_1^{(-)}$, where the area under the curve between $X_1^{(+)}$ and $C$ (denoted $B_1$) is equal to the area under the curve between $C$ and $X_1^{(-)}$. The second method compares $X_1^{(+)}$ and $Y_1^{(+)}$, which are equidistant from $C$, but the areas under the curve between these points and $C$ are different.


\section{Simulations} \label{sec:simulation}
\begin{figure}[htbp]
    \centering
\begin{tikzpicture}
    \begin{axis}[
        no markers, 
        domain=-2:5, 
        samples=200, 
        axis lines=left, 
        xlabel=$x$, 
        ylabel=$f(x)$,
        xlabel style={at={(ticklabel* cs:1)}, anchor=north west},
        ylabel style={at={(ticklabel* cs:1)}, anchor=south west},
        height=6cm, 
        width=10cm,
        xtick={0.2, 1, 2.1, 3.0}, 
        xticklabels={LB, 0, $\beta$, UB},
        ytick=\empty,
        enlargelimits=upper, 
        clip=false
    ]

      \addplot [very thick, black] {exp(-(x-1)^2/2) / sqrt(2*pi)};

      \draw [ultra thick, blue!80!black] (axis cs:0.2, 0) -- (axis cs:3.0, 0);

      \draw [dotted, thick, black] (axis cs:1, 0) -- (axis cs:1, 0.3989);

      \draw [dotted, thick, black] (axis cs:2.1, 0) -- (axis cs:2.1, 0.217);

      \draw [dashed, thick, red] (axis cs:0.2, 0.289) -- (axis cs:0.2, 0);

      \draw [dashed, thick, red] (axis cs:3.0, 0.054) -- (axis cs:3.0, 0);
    \end{axis}
\end{tikzpicture}
    \caption{Simulation exercise}
    \label{fig:simulation}
\end{figure}
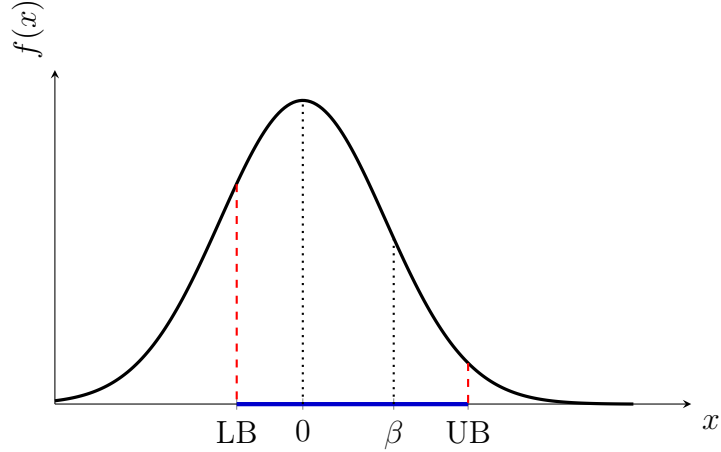

To gain an understanding of how the proposed testing methodology performs, we conducted a simulation exercise. Specifically, we started by generating a perfectly symmetric standard Normal (0,1) and we assumed that the cutoff is $C=0$. We then set a new value of the cutoff, $C+\beta$, with $\beta>0$ and computed the BW extremes (UB and LB). This step generates a data-driven asymmetric interval around $C=0$, which is then used in the subsequent step. Considering the true cutoff value $C=0$, in such an uneven BW, we compute the testing indices to check whether they could capture the asymmetry and signal its direction. One question that may arise at this point is: why not look at the symmetry indices when $C=\beta$? The result in this case would be obvious, the indices clearly point to asymmetry. However, we want to verify the potential of the indices to signal asymmetry even with a symmetric distribution, by considering an asymmetric interval that makes the distribution asymmetric \textit{de facto}. This is a rather finer exercise. The following pseudo-code provides a schematic breakdown of the simulation exercise for $Q=1,000$ repetitions, and Figure \ref{fig:simulation} illustrates the process visually. Notice that the threshold $\alpha$ is used only for the identification of BW. Moreover, we denote by $N$ the highest number of iterations to compute $n_\alpha^\star$, defining the optimal BW.

\begin{enumerate}
\item Set $C=0$;
\item Initialize the other parameters of the simulation $\alpha$, $\beta$, $N$;
\item Set seed;
\item Set $q=1$;
    \item generate a perfectly symmetric running variable from a N(0,1), such that the absolute values of the negative values are equal to the positive values;
    \item shift the cutoff from $C$ to $C+\beta$;
    \item compute the BW as in \eqref{nstar} and retrieve $UB$ and $LB$;
    \item within the interval $\left[ LB, UB \right]$ compute $X_{i}^{(+)}$, $X_i^{(-)}$, $Y_{i}^{(+)}$, $Y_{i}^{(-)}$ for $i=1, \dots, 9$ and apply eq. \eqref{eq:test} and \eqref{eq:MADs};
    \item Set $q=q+1$. If $q=Q$, stop. Else, go to step 3.
\end{enumerate}

Since the cutoff has been shifted rightward, we expect $|LB|<UB$, implying that the interval on the LHS of the cutoff point is shorter than the corresponding interval on the RHS, thereby endogenously generating an asymmetric BW around zero. In principle, the exercise can be repeated for different values of $\beta$ as long as $LB<0$; namely, for sufficiently small shifts. \\ Given this setup, we expect $MAD^{(-)}$ to point toward symmetry, and conversely for $MAD^{(+)}$. This follows from the fact that the values of $X_{i}^{(+)}$ are expected to be larger than $X_{i}^{(-)}$ (in absolute terms), so that when reflecting $X_{i}^{(+)}$'s values on the LHS (i.e., when generating $Y_i^{(+)}$), some or even all may exceed $|LB|$. In the presence of a marked asymmetry of the BW, it can even happen that $Y_1^{(+)}$ is to the left of LB, $|Y_1^{(+)}|>|LB|$. In such a situation, all the probability mass on the LHS is contained within $LB$ and the cutoff, and the value of $MAD^{(+)}$ will be high.\footnote{If this situation applies $MAD^{(+)}=\frac{1}{9}\left[(1-0.301)+(1-0.301)\right]=0.155333$.} \\  Summing up: we expect the following three results from the simulation exercise: (i) $|LB|<UB$, (ii) $MAD^{(-)}$ taking low values (iii) $MAD^{(+)}$ taking high values. \\
Therefore, in accordance with the discussion below Formula \eqref{eq:MADs}, we can introduce the concept of `partial symmetry' or `left(right)-symmetry.' This apparent oxymoron describes a situation in which only one side of the distribution conforms to BL. Such a finding can be quite informative depending on the specific economic context under scrutiny. For instance, consider a scenario where units scoring above the cutoff are entitled to a benefit. We would expect significantly more units on the RHS, resulting in $UB \gg |LB|$ and a disproportionately high $MAD^{(+)}$.

\subsection{Results} 
\label{sec:sim_perfect}
This exercise can be viewed as an \textit{in vitro} experiment aimed at studying the behavior of the test proposed by thoroughly isolating the causes of potential success or failure. To enhance the readability of the simulation results, and without loss of generality, we adopted a single critical value for each metric. 

The threshold $\lambda$ used for the case of Formula \eqref{eq:MADs} is selected in accordance with \cite{nigrini}.
Indeed, in close analogy to the popular use of critical $p$-values at $1\%$, $5\%$ and $10\%$ for rejection of the null hypothesis, \cite{nigrini} identifies some variation ranges of the value of MAD to have close/acceptable/marginal conformity with BL. Notably, the threshold values for $\alpha$ are $0.006$, $0.012$ and $0.015$ so that MADs below $0.006$ identify \textit{close} conformity, values between $0.006$ and $0.012$ \textit{acceptable} conformity and between $0.012$ and $0.015$ \textit{marginal} conformity.\footnote{See \cite{cerqueti2021data} Table 2 for a correspondence between the MAD thresholds and other criteria elaborated in the literature.} While alternative critical values can be chosen, adopting the threshold established by \cite{nigrini} provides an additional strength to our approach.

Since only $0.5\%$ of the cases for the $MAD^{(-)}$ fall between the close and marginal acceptance thresholds (i.e., between $0.006$ and $0.015$), we used the close conformity threshold in our experiments. Similarly, for $MAD^{(+)}$, even if this proportion is slightly higher. \\
Very briefly, the results can be summarized in the following three points: 
\begin{itemize}
    \item $|LB| < UB$ \hspace{90pt} $98.20\%$
    \item $MAD^{(-)}$ points to symmetry $99.40\%$ ($< 0.006$)
    \item $MAD^{(+)}$ points to asymmetry $95.30\%$ ($> 0.006$) 
\end{itemize}

 As an initial result, the average value of the ratio $\text{LB}/\text{UB}$  is around $94\%$  and the mean $MAD^{(-)}$ value of $0.0008$ falls well below the critical threshold for close conformity. In $98.2\%$ of the cases $|LB| < UB$. \\  A breakdown of the results by BW length is reported in Tables \ref{tab:LB_MAD-} and \ref{tab:LB_MAD+} for $MAD^{(-)}$ and $MAD^{(+)}$, respectively. Table \ref{tab:LB_MAD+} reveals that in all cases where $|LB| < UB$, $MAD^{(-)}$ indicates symmetry (lower-right quadrant). Furthermore, the $MAD^{(-)}$ remains robust even when interval proportions shift -- specifically when  $|LB| \geq UB$. Even under these conditions, the measure correctly identifies symmetry in $1.2\%$ of cases (upper-right quadrant), bringing the total symmetry detection to $99.4\%$.

\begin{table}[!h]
\centering
\caption{Cross-tabulation: BW and $MAD^{(-)}$ - perfect symmetry}
\begin{tabular}{|l | l|c|c|}
\hline
  &   & \multicolumn{2}{c|}{$MAD^{(-)} < 0.006$} \\ \hline
  &   & NO & YES \\ \hline

\multirow{2}{*}{\makecell[c]{$|LB|<UB$}}
  & NO  & 0.60 & 1.20 \\ \cline{2-4}
  & YES & 0.00 & 98.2 \\ \hline
\end{tabular} 
\label{tab:LB_MAD-}

\begin{footnotesize}
\vspace{5pt}
\centering
\textbf{Note:} the table reports percentages out of $1,000$ repetitions
\end{footnotesize}            
\end{table}

Regarding $MAD^{(+)}$, a converse pattern is observed. Its mean value of $0.0254$ stands above the threshold for close conformity. Table \ref{tab:LB_MAD+} provides a cross-tabulation akin to that of Table \ref{tab:LB_MAD-}. The most significant finding is in the lower-left quadrant: whenever $|LB| < UB$, $MAD^{(+)}$ indicates asymmetry in $95.30\%$ of cases, with the remaining $4.7\%$ pointing to close conformity.

\begin{table}[!h]
\centering
\caption{Cross-tabulation: BW and $MAD^{(+)}$ - perfect symmetry}
\begin{tabular}{|l | l|c|c|}
\hline
  &   & \multicolumn{2}{c|}{$MAD^{(+)} < \textbf{0.006}$} \\ \hline
  &   & NO & YES \\ \hline

\multirow{2}{*}{\makecell[c]{$|LB|<UB$}}
  & NO  & 0.00  & 1.80  \\ \cline{2-4}
  & YES & 95.30 & 2.90 \\ \hline
\end{tabular} 
\label{tab:LB_MAD+}

\begin{footnotesize}
\vspace{5pt}
\centering
\textbf{Note:} the table reports percentages out of $1,000$ repetitions
\end{footnotesize}            
\end{table}
To evaluate the quality of the results, additional reference values beyond Nigrini's thresholds have recently been proposed by \cite{cano2025divergence}, who tabulated the critical values of the MAD test. In particular, $0.3216$, $0.3492$, and $0.4046$ correspond to the $90$th, $95$th, and $99$th percentiles, respectively. However, the MAD score exhibits a mechanical negative dependence on the number of observations $M$, leading to different critical values for each value of $M$. This dependence can be accounted for by multiplying the MAD by $\sqrt{M}$. Equivalently, one can divide the percentiles by $\sqrt{M}$ \citep{barney2016moderating, de2023study, nigrini2015persistent}. The principal drawback to this approach is that, with a sufficiently large $M$, even small departures from the tabulated MAD will result in a finding of no conformity; even more oddly, a large MAD with a small $M$ is more likely to lead to a conclusion of conformity than a low MAD with a larger $M$. In order not to leave this approach unexplored, we have counted how many times the MAD values are below $\frac{P_{95}}{\sqrt{M}}$ -- being $P_{95}$ the $95$th percentile -- finding that $MAD^{(-)}$ does not reject conformity to BL $100\%$ of the time, and $MAD^{(+)}$ rejects it $99.88\%$ of the time. \\

A further complementary measure of non-conformity to BL is provided by the Equivalent Contamination Proportion (ECP). This represents the proportion of non-BL conforming observations within an otherwise BL-conforming sample that would result in an expected MAD value equal to the one observed in the actual data, see \cite{cano2025divergence} and \cite{cano2025much}. This measure presents the advantage of being independent of sample size and offers consistent results across different divergence statistics. When applied to our simulation this methodology returns an ECP equal to $0.0\%$ and $40.84\%$ for $MAD^{(-)}$ and $MAD^{(+)}$, respectively. \\
As far as the tolerance threshold approach is considered, eq. \eqref{eq:test}, we get $D(\tilde{\mathcal{X}}^{(-)}, \tilde{\mathcal{X}}^{(+)})=0.0737$.

Finally, for the sake of completeness, we have supplemented the simulation results with McCrary-type tests. A comparison between the two methodologies is not straightforward, as the McCrary-type test only allows for symmetric BW. This implies that repeating the experiment on the perfectly symmetric generated distribution would be a redundant exercise, resulting in the acceptance of the null hypothesis of symmetry $100\%$ of the time with the highest possible p-value. \\ Consequently, to ensure an informative comparison, we proceeded as follows: first, we allowed the \texttt{rddensity} routine to endogenously compute the symmetric BW around the cutoff $C=0$. Second, we rescaled the BW by applying the average proportions obtained from the simulation (i.e., $LB/UB$). Third, we replicated the McCrary test within this interval. The result is a non-rejection of the null hypothesis of symmetry ($p\text{-value} = 0.850$). \\ As a further robustness check, we repeated the test using a narrower BW, determined by the average $LB$ and $UB$ from the simulation, and obtained a qualitatively unaltered result ($p\text{-value} = 0.986$). \\
Overall, these results highlight the complementary nature of the two approaches. While the McCrary-type test provides a useful baseline by confirming global symmetry, our proposed methodology offers additional layers of information by successfully identifying specific rejections of right-symmetry. This suggests that our test serves as a more granular diagnostic tool, capable of detecting nuanced distributional asymmetries that remain overlooked by traditional tests.

\section{Applications}
\label{sec:applications}
We illustrate our methodology using two popular datasets that have already been used for similar purposes. In the first example we used the dataset from \cite{meyersson2014islamic} who studied the causal effect of Islamic political representation winning municipal elections in Turkey on women's highest educational degree. The score in this RDD is the margin of victory of the largest Islamic party in the municipality.
The second example is taken from \cite{londono2020upstream}, in which the authors examine how financial aid influences post-secondary enrollment, college choice, and student composition. Their analysis is based on a large-scale program in Colombia that provides high-achieving, low-income students with access to high-quality colleges. 
Eligibility for the program requires students to exceed a cutoff in a national standardized high school exit exam (SABER) and to fall below a threshold in the household wealth index (SISBEN). The empirical analysis focuses on the discontinuity generated by the SISBEN cutoff, treating the achievement requirement as an additional eligibility condition. The choice of these datasets is motivated by evidence of continuity of the running variable around the cutoff, as documented by the McCrary test (see the original papers). However, the distributions of the variables considered are asymmetric, highlighting a key difference between the present paper and the existing contributions.
\subsection{Application 1: the \cite{meyersson2014islamic} dataset}
 \label{sec:case1}
\begin{figure}[h]
    \centering
        \includegraphics[scale = 0.45]{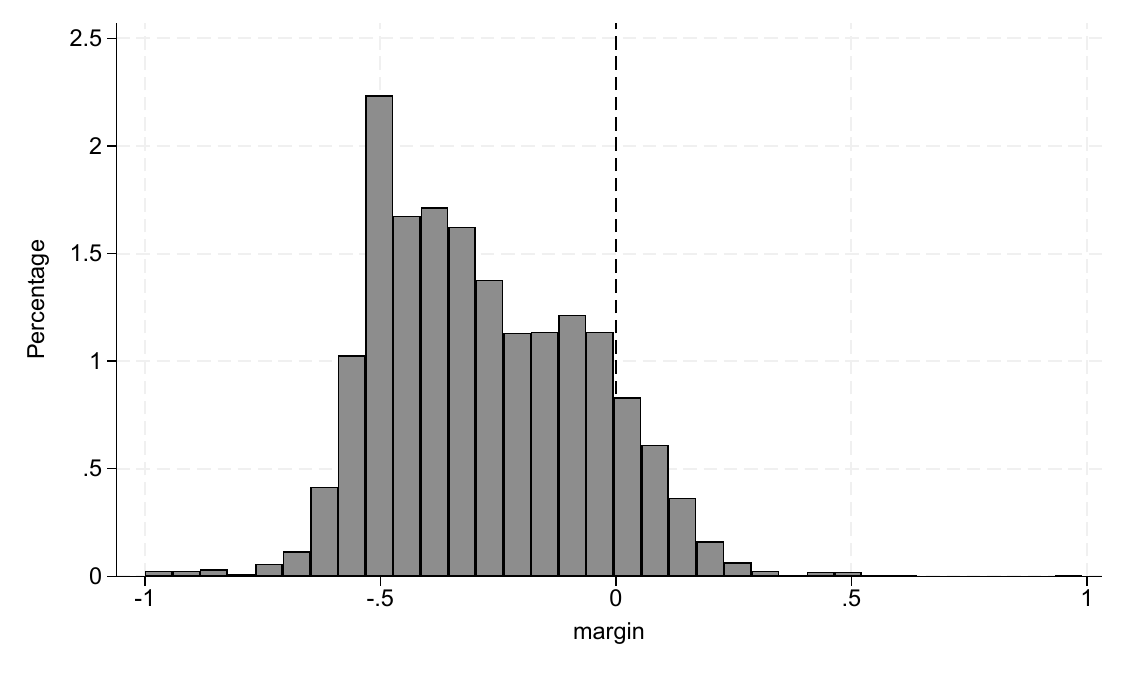}
    \caption{Density plot of the running variable for the \cite{meyersson2014islamic} case.}
    \label{fig:meyerson}
\end{figure}

Figure \ref{fig:meyerson} plots the empirical density of the normalized running variable which ranges from $-1$ up to $0.9905$ with an average value of $-0.281$ and a median around $-0.314$. 

The iterative procedure for BW selection has been repeated ten times ($N=10$) and the MADs are reported in Table \ref{tab:BW1}.
\begin{table}[ht]
\begin{center}
\caption{BW choice, the case of \cite{meyersson2014islamic} }
    \begin{tabular}{l c c}
        \textbf{Iterate} & \textbf{MAD}$^{(+)}$ & \textbf{MAD}$^{(-)}$  \\ \hline
        1 & 0.0028 & 0.0002  \\ 
        2 & 0.0031 & 0.0004  \\ 
        3 & 0.0038 & 0.0012  \\ 
        4 & \textbf{\textcolor{red}{0.0059}} &  \textbf{\textcolor{red}{0.0023}}  \\ 
        5 & 0.0114 & 0.0031  \\ 
        6 & 0.0200 & 0.0041  \\ 
        7 & 0.0616 & 0.0125  \\ 
        8 & 0.0913 & 0.0220  \\ 
        9 & 0.1338 & 0.0330  \\ 
        10 & 0.1553 & 0.0643  \\ \hline
    \end{tabular}
     \label{tab:BW1}
    \end{center}
\begin{footnotesize}
\textbf{Note}: the values of $MAD^{(+)}$ and $MAD^{(-)}$ are computed using the iterative procedure of Section \ref{sec:BW}, from iterate 1 to 10. 
\end{footnotesize}            
\end{table}
 For the values to the RHS of the cutoff, the iterate 4 identifies both $MAD^{(-)}$ and $MAD^{(+)}$ in \textit{close conformity} to the BL, with $MAD^{(-)}_4=0.0023<0.006$ and $MAD^{(+)}_4=0.0059<0.006$. It follows that the optimal BW is: $\left[-X_{4,9}, X_{4,9}\right]$, corresponding to $\left[-0.1138, 0.0274\right]$ . \\ 
We can now proceed to apply the two types of tests. In the tolerance threshold approach, we set $M=0.1138$ and $C=0$ so that \\
\begin{center}
$\tilde{X}^{(-)}_j=\frac{0-X^{(-)}_{4,j}}{\Gamma} \qquad $ and \qquad $\tilde{X}^{(+)}_j=\frac{X^{(+)}_{4,j}-0}{\Gamma} \qquad \forall \ \ j=1, \dots, 9$. \\    
\end{center}

Accordingly, $D(\tilde{\mathcal{X}}^{(-)}, \tilde{\mathcal{X}}^{(+)})$ is computed as:
\begin{equation}
D(\tilde{\mathcal{X}}^{(-)}, \tilde{\mathcal{X}}^{(+)})=\frac{1}{9} \sqrt{\sum_{j=1}^9 |\tilde{X}^{(-)}_j-\tilde{X}^{(+)}_j|} = 0.2695.
\end{equation}
In the MAD approach, we set $Y_j^{(+)}=X_{6,j}^{(-)}$ and $Y_j^{(-)}=X_{6,j}^{(+)}$ for $j=1, \ldots, 9$, so that equations \eqref{eq:prop+} and \eqref{eq:prop-} can be operationalized as the sum of the score values within the intervals defined by $Y_j^{(+)}$ and $Y_j^{(-)}$, respectively. This yields $MAD^{(+)}=0.0502$ and $MAD^{(-)}=0.0254$. Interestingly, $MAD^{(+)} > MAD^{(-)}$, indicating that the right-asymmetry is greater than the left and that the eligibility is concentrated on the RHS. When these indices are rescaled to account for the sample size, we obtain $\sqrt{M} \cdot MAD^{(+)}=0.4166$ and $\sqrt{M}\cdot MAD^{(-)}=0.4760$, both of which are slightly higher than the $99$th percentile $P_{99}$ threshold (0.4046). Finally, the ECP indicates that the tabulated MAD values correspond to a BL-conforming distribution, with $68.87\%$ of corrupted data on the RHS and $37.42\%$ on the LHS. 

\subsection{Application 2: the \cite{londono2020upstream} dataset}
 \label{sec:case2}
\begin{figure}[ht]
    \centering
    \includegraphics[width=0.6\textwidth]{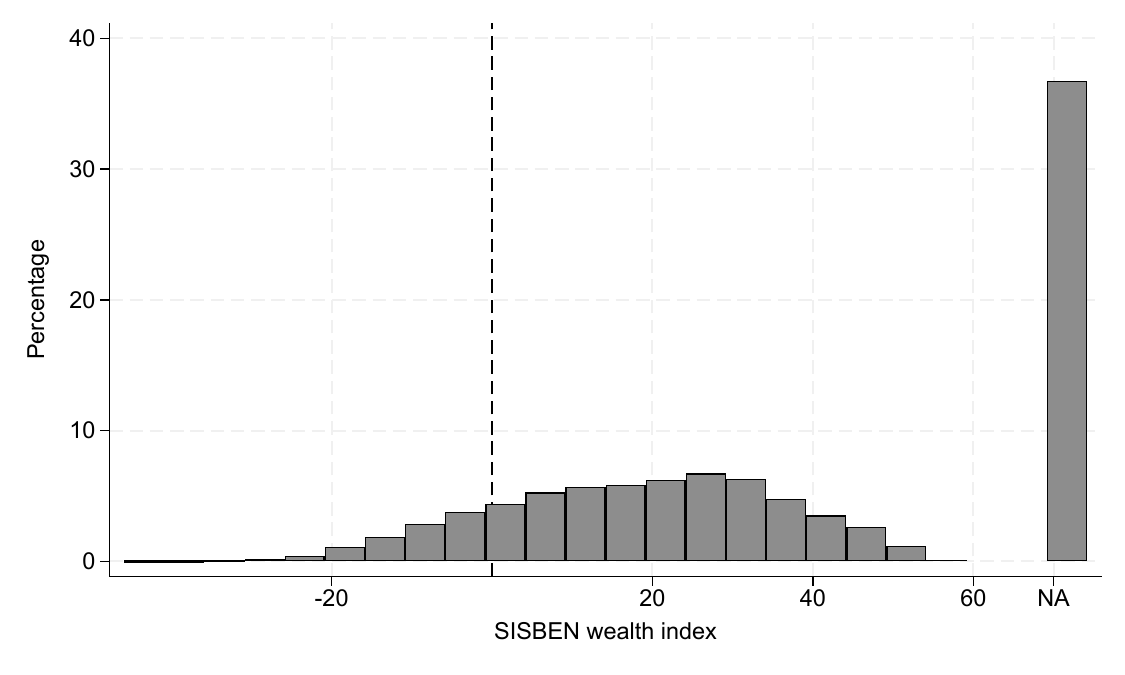} 
    \caption{Density plot of the SISBEN score running variable (Londo\~no V\'elez et al., 2020)}
    \label{fig:londono_BNORM}
\end{figure}

Figure \ref{fig:londono_BNORM} plots the empirical density of the centered running variable for the \cite{londono2020upstream} case. Specifically, the distance to the eligibility cutoffs in terms of the households' wealth index, the SISBEN score. It is immediately apparent that the distribution of the index is slightly left-skewed. The index ranges from $-45.85$ to $56.40$, with a mean of $17.96$ and a median of $19.12$.

The results of the iterative procedure for the BW selection are reported in Table \ref{tab:BW3}. In this case, we have $MAD_6^{(+)}=0.0016$ as being the last consistent with \textit{close conformity} for positive values of the score, and the corresponding $MAD_6^{(-)}=0.0034$. Thus, we have the boundaries of the BW $\left[-X_{6,9}^{(-)}, X_{6,9}^{(+)}\right]$ corresponding to  $\left[-0.92 , 1.91\right]$. \textit{Mutatis mutandis}, going through the same steps as for the Meyersson case, we obtain $D(\tilde{\mathcal{X}}^{(-)}, \tilde{\mathcal{X}}^{(+)})$ = 0.2254, $MAD^{(+)}=0.0154$ and $MAD^{(-)}= 0.0257$.  
 \begin{table}[h]
\begin{center}
\caption{BW choice for the SISBEN score, \cite{londono2020upstream} }
    \begin{tabular}{l c c}
        \textbf{Iterate} & \textbf{MAD}$^{(+)}$ & \textbf{MAD}$^{(-)}$  \\ \hline
        1 & 0.0001 & 0.0003  \\ 
        2 & 0.0002 & 0.0011  \\ 
        3 & 0.0004 & 0.0037  \\ 
        4 & 0.0009 & 0.0014  \\ 
        5 & 0.0005 & 0.0058  \\ 
        6 & \textcolor{red}{\textbf{0.0016}} & \textcolor{red}{\textbf{0.0034}}  \\ 
        7 & 0.0068 & 0.0053  \\ 
        8 & 0.0067 & 0.0145  \\ 
        9 & 0.0092 & 0.0456  \\ 
        10 & 0.0325 & 0.0731  \\ \hline
    \end{tabular}
     \label{tab:BW3}
    \end{center}
\begin{footnotesize}
\textbf{Note}: the values of $MAD^{(+)}$ and $MAD^{(-)}$ are computed using the iterative procedure of Section \ref{sec:BW}, from iterate 1 to 10. 
\end{footnotesize}            
\end{table}
It is worth noticing that in this case eligible individuals are located on the LHS of the cutoff and that we computed a $MAD^{(-)} > MAD^{(+)}$, indicating that the left-asymmetry is greater than the right. In addition,  $\sqrt{M}\cdot MAD^{(+)}=0.4354$ and $\sqrt{M} \cdot MAD^{(-)}=0.5401$, while the ECP is $22.05\%$ and $39.25\%$ for right and left-asymmetry, respectively.

\section{Concluding remarks}
\label{sec:conclusion}
The paper proposes a new methodology for the identification of possible data manipulations in a RDD context. Based on the BL, we first propose an iterative procedure to select a BW within which to carry out the tests. The test is broken down into two mechanisms. The first compares threshold values of the running variable while keeping the density constant, and the second compares densities while keeping the threshold values constant. This methodology can be considered complementary to the popular McCrary-type tests and offers some clear advantages. First, it does not require researcher prespecified parameter settings. Second, it overcomes the limitations of the law itself. Third, the BW within which to carry out the test is automatically suggested by BL, again avoiding discretionary choices regarding the optimum data-driven method to be chosen. The deviation of only one of the two MADs from the critical values may suggest further information on the asymmetry properties of the manipulation. In this case, we suggest considering only the MAD on the side where units are incentivized to place their score. Finally, two empirical applications suggest that our procedure is more prudential than a McCrary-type test in detecting asymmetries; therefore, one cannot refrain from employing it as a necessary complement. For instance, if the canonical McCrary test suggests manipulation, examining the different values of $MAD^{(+)}$ and $MAD^{(-)}$ may be highly informative about the source of the deviation. If this deviation originates solely from the side where units have no incentive to place their score, a more lenient final verdict may be appropriate-even if asymmetry is still present. Similarly, our proposed test can support empirical decisions when the traditional benchmark suggests rejecting the null at the $10\%$ level but not at $5\%$. In sum, while the McCrary test remains the industry standard, its reliance on symmetric bandwidths inherently limits its diagnostic power, occasionally failing to reject the null even when underlying asymmetries are present. In contrast, our approach offers a more nuanced decomposition by explicitly distinguishing between left- and right-symmetry. By uncovering a rejection of one-sided symmetry where traditional frameworks remain uninformative, this new methodology provides researchers with a more granular lens through which to evaluate density behavior, ensuring a more robust and comprehensive validation of the identifying assumptions in RDD settings.
\\ 
At the same time, some limitations of the proposed approach call for further investigation. In particular, additional research is needed to establish clearer operational thresholds for accepting or rejecting the null of symmetry as a function of the observed MAD. A key challenge in this respect is the dependence of the MAD on sample size, which complicates the definition of universal or easily interpretable critical regions. Relatedly, future work could aim at deriving or approximating critical values for alternative summary measures, such as the ECP, in order to provide complementary and potentially more scale-invariant decision criteria. Addressing these issues would further enhance the interpretability and practical applicability of the proposed testing framework.

\newpage
\bibliographystyle{plainnat}
\bibliography{bibl.bib}

\end{document}